\documentclass[preprints]{Definitions/mdpi} 
\firstpage{1} 
\pubvolume{1}
\issuenum{1}
\articlenumber{0}
\pubyear{2026}
\copyrightyear{2026}
\datereceived{ } 
\daterevised{ } % Comment out if no revised date
\dateaccepted{ } 
\datepublished{ } 
\preto{\abstractkeywords}{\nolinenumbers}

\Title{Primordial Black Holes Formation Beyond the Standard Cosmic QCD Transition}

\Author{Ma\"el Gonin$^{1,2,*}$\orcidA{} 
and Oleksii Ivanytskyi$^{3,4,5}$\orcidB{}
and David Blaschke$^{3,4,5}$\orcidC{} 
and G\"unther Hasinger$^{1,2}$\orcidD{}
}

\AuthorNames{Ma\"el Gonin and Oleksii Ivanytskyi and David Blaschke and G\"unther Hasinger
}

\address{%
$^{1}$ \quad Deutsches Zentrum f\"ur Astrophysik (DZA), Postplatz 1, 02826 G\"orlitz, Germany\\
$^{2}$ \quad Institute of Nuclear and Particle Physics, Technical University Dresden, 01062 Dresden, Germany\\
$^{3}$ \quad Institute of Theoretical Physics, University of Wroclaw, Max Born Pl. 9, 50-204 Wroclaw, Poland\\%; david.blaschke@uwr.edu.pl\\
$^{4}$ \quad Helmholtz-Zentrum Dresden-Rossendorf (HZDR), Bautzner Landstrasse 400, 01328 Dresden, Germany\\
$^{5}$ \quad Center for Advanced Systems Understanding (CASUS), Untermarkt 20, 02826 G\"orlitz, Germany%; e-mail@e-mail.com}
}

\corres{Correspondence: mael.gonin@dzastro.de
}

\abstract{
We review the role of primordial black holes (PBHs) for illuminating the dark ages of the cosmological evolution and as dark matter (DM) candidates. We elucidate the role of phase transitions for primordial black hole formation in the early Universe and focus our attention on the cosmological QCD phase transition within a recent microscopical model. We explore the impact of physics beyond the Standard Model (SM) on the cosmic equation of state and the probability distribution for the formation of PBHs which serve as candidates for DM and contribute to present-day binary black-hole merger events.
}


\begin{document}

\section{Introduction}

The search for sub-solar mass mergers is of particular interest for cosmology and astrophysics, as such compact objects hints for new physics \cite{LVK:2022ydq,2025arXiv251119911T,2026arXiv260212115K}. On November 12, 2025, the LVK collaboration reported a candidate sub-solar mass gravitational wave merger event \cite{LISACosmologyWorkingGroup:2023njw,2025GCN.42724....1S}. This single candidate, if confirmed, could be a direct evidence that PBHs populate the sub-solar mass window — a region inaccessible to standard stellar evolution and therefore a unique fingerprint of early Universe physics \cite{2026arXiv260106024C}. However, another interpretation of this event in terms of a subsolar mass neutron star merger have been discussed as well \cite{Kasliwal:2025keb}.
PBHs form in the radiation-dominated era when sufficiently large overdensities cross the particle horizon \cite{Carr:1974nx}. Unlike other relics, they carry direct information on the Universe before Big Bang Nucleosynthesis (BBN): their mass spectrum is a fossil of the cosmic equation of state (EoS) at their formation \cite{Carr:1975qj,Byrnes:2018clq,Musco:2023dak,Gonin:2025uvc}. The PBH distribution from the thermal history is constrained by many observations \cite{Carr:2020xqk,Carr:2020gox,Carr:2019kxo,Hasinger:2020ptw,LISACosmologyWorkingGroup:2023njw,2026arXiv260106024C}. Any modification of the thermal history therefore leaves a direct imprint on the PBH mass distribution, and mitigates the constraints. 

The Big Bang scenario describes the early Universe as an extremely dense plasma of fundamental particles, cooling and expanding on a radiation-dominated background. As it cools, a succession of phase transitions (PT) occurs. It is of primary importance to map the nuclear matter phase diagram, to know how the Universe evolved through this period and what kind of PT remnants can be found in observations \cite{Borsanyi:2016ksw,Pandav:2025sqo,Hindmarsh:2020hop,Guenther:2020jwe,Guenther:2022wcr,Lu:2023msn}.
While the SM predicts smooth crossovers for both the QCD and electroweak (EW) transitions, the motivation for at least one strong 1st order PT remains compelling: a strong 1st order PT can leave imprints in the Stochastic Gravitational Wave Background (SGWB) through bubble nucleation \cite{Hindmarsh:2020hop,Correia:2025qif}, and its order has direct consequences for baryogenesis \cite{Bodeker:2020ghk, 2025arXiv250809989V}. The Sakharov conditions \cite{Sakharov:1967dj} require CP violation, departure from thermal equilibrium, and baryon number violation — conditions a 1st order PT naturally satisfies. The smooth EW crossover suggested by the observed Higgs mass $m_H=125~\rm GeV$ \cite{ParticleDataGroup:2022pth} seems to disfavor EW baryogenesis in the SM \cite{Kajantie:1996mn,Kajantie:1996qd}, yet the research for beyond standard model (BSM) mechanisms remains very active \cite{Boeckel:2010bey,Boeckel:2011yj,Iso:2017uuu,Khlopov:2021xnw}.

Asymmetries in the lepton sector could also be powering 1st order transitions. Schwarz \& Stuke \cite{Schwarz:2009ii} showed that an accurate description of the early Universe requires accounting for lepton flavor asymmetries, which generate cosmic trajectories in the QCD phase diagram. A large primordial lepton asymmetry of the Universe (LAU) can suppress sphaleron rates, reconciling a large LAU with the observed small baryon asymmetry (BAU) \cite{March-Russell:1999hpw,Barenboim:2017dfq}, and can trigger a 1st order QCD transition with a distinct SGWB signature \cite{Gao:2021nwz,Gao:2023djs,Gao:2024fhm}. Crucially, large lepton asymmetries before BBN are observationally allowed, as neutrino flavor oscillations starting at $T\sim 10 ~\rm~MeV$ relax them to values consistent with CMB and BBN constraints \cite{Barenboim:2016shh, Froustey:2024mgf}. B\"odeker et al. demonstrated that these asymmetries reshape the PBH spectrum in a way potentially consistent with LVK observations \cite{Bodeker:2020stj}.

In this work we extend this framework using a microscopic QCD model \cite{Blaschke:2023pqd}, incorporating BAU and LAU through state of the art Taylor-expanded susceptibilities up to 4th order, following \cite{Formaggio:2025nde}. We compute cosmic trajectories and the EoS across a range of baryon and lepton asymmetry scenarios and translate the results into PBH mass distributions, discussing their implications for gravitational wave and microlensing observations.

%%%%%%%%%%%%%%%%%%%%%%%%%%%%%%%%%%%%%%%%%%%%%%%%%
%%%%%%%%%%%%%%%%%%%%%%%%%%%%%%%%%%%%%%%%%%%%%%%%%

\section{The thermal history after the Big Bang}
\label{sec:EoS}
In this section we develop on the cosmic EoS accross different transitions. We start by presenting the standard model picture, before discussing effects of Beyond-SM physics.

\subsection{The Standard Model}\label{subsec:SM}

The cosmological equation of state is described through SM interactions, assuming that the laws of physics are rigorously the same as those observed on Earth at present time. The Big Bang scenario describes the early Universe as an extremely dense plasma of fundamental particles in thermal equilibrium on an expanding background. The rate of particle interactions being much larger than the Hubble expansion rate \cite{Rafelski:2024fej}, the primordial plasma can be described solely through thermodynamic equilibrium. In its early stage the "cosmic soup" is radiation dominated\footnote{This study focuses on the radiation dominated era, later times are outside the scope of the current study.} and cools down with time according to \cite{Husdal:2016haj}:
\begin{equation}
    t=\sqrt{\frac{90\hbar^3c^5}{32\pi^3Gg_{e}(T)}}(k_BT^{-2}) \approx \frac{2.4~{\rm s}}{\sqrt{g_e(T)}}T_{MeV}^{-2}
\end{equation}
where $g_\varepsilon = \varepsilon \frac{30}{T^4 \pi^2}$ is the effective number of relativistic degrees of freedom for the energy density $\varepsilon$.
As it cools down, transitions to different phases of matter occur; let us make a brief summary. After the Planck time $t_{\rm Planck}=10^{-43} \rm s$ corresponding to a temperature $T_{\rm Planck}\sim 10^{19} \rm GeV$, when all four fundamental interactions are unified, the Universe is minuscule, and quantum fluctuations are non-negligible. After $t_{\rm Planck}$, gravity decouples and the Grand Unification Era begins. It lasts until the strong interaction decouples at $t_{\rm strong}=10^{-36}$ s with $T_{\rm strong} = 10^{16}\rm~GeV$, which triggers the inflation era. 
According to Guth's inflationary scenario \cite{Guth:1980zm}, if the Universe undergoes a 1st order phase transition with sufficient supercooling during this early epoch, it can enter a period of exponential expansion driven by the energy density of a false vacuum state. 
This inflationary phase addresses two fundamental problems of the standard Big Bang model: the horizon problem and the flatness problem. 
The exponential expansion during inflation stretches quantum fluctuations, filling the Universe with patches of inhomogeneity in energy density and curvature.

The decay of the inflation field marks the start of the quark-gluon plasma (QGP) phase. The deconfined quarks and ambient gluons provide the bulk of the contribution to the thermodynamics until the temperature drops down to $T_c\approx156.5 ~\rm~MeV$ \cite{HotQCD:2018pds},
%\textbf{(@OLEKSII \& DAVID can you add references for the pseudo critical QCD temperature ?)} 
when a smooth crossover to the hadron resonance gas (HRG) occurs as quarks get confined into hadrons that (except protons) subsequently decay. 
Such energy regimes can be explored experimentally through large scale projects such as the LHC \cite{Pandav:2025sqo}. In heavy ion collisions the fireball formed can be similar to the very early Universe in terms of temperature regime but it is significantly different when looking at the baryon's net number density and chemical potential. Also the lifetime of the experimental fireball is much smaller than the Big Bang time scale. Therefore one cannot straightforwardly apply LHC results to the Big Bang model. It is of primary importance to properly map the QCD phase diagram for cosmology and particle physics experiments, see \cite{Guenther:2020jwe,Guenther:2022wcr} for a recent overview.
The preferred way to explore the nuclear matter phase diagram in the Big Bang regime is through lattice QCD simulations. In this study we rather use the generalized
Beth-Uhlenbeck (GBU) approach to the EOS of strongly
interacting matter, which we will refer to as the microscopical model. 

According to the GBU approach, hadrons appear as correlations in the strong interaction among quarks which are described by phase shifts with
a characteristic jump by $\pi$ for a bound state and a smooth tail describing the continuum of scattering states. These phase shifts in different hadronic
channels are subject to in-medium modifications such as the chiral symmetry restoration ($\chi$SR) and bound state dissociation (Mott effect).
When one reverses the cosmic evolution so that the temperature increases and passes $T_c$, the chiral condensate melts and the quark masses drop towards their current-quark mass values
in the $\chi$SR crossover transition. As a consequence the continuum edges for two-, three-, and multiquark states are dramatically lowered and "eat up"
the hadronic bound states which upon further temperature increase disappear as fading resonances in the continuum of few-quark scattering states.
At the same time, when the quarks loose most of their mass upon $\chi$SR, they appear as dominant degrees of freedom together with the gluons which are set free
by the simultaneous breaking of the center symmetry of the SU(3) color charge of QCD. This aspect of the QCD transition is captured in the behaviour of the traced Polyakov loop and its potential. With the appearance of deconfined quarks and gluons in their QGP state, also  perturbative quark-gluon scattering processes play an important role and have been included to the microscopic model which provides an excellent description of QCD thermodynamics in accordance with the ab-initio lattice QCD simulations. For a detailed description and further results of the approach, see \cite{Blaschke:2023pqd}.
An advantage of the microscopic approach over that of the lattice QCD simulation is its flexibility when applications to cosmology are considered. We can straightforwardly extend the temperature range of the calculation, add photons, leptons, further quark flavors and hadron species as well as extend the application to nonzero chemical potentials of baryon number and different flavors. 
%\textbf{(@Oleksii \& David maybe add a small paragraph ?)}

Once the pressure $P$ as thermodynamic potential of the grand canonical ensemble is known, thermodynamic relations can be used to obtain the entropy density $s$,  the number densities $n_i$ and the energy density $\varepsilon$:
\begin{subequations}
\label{eq:thermo}
\begin{flalign}
    s &= \frac{dP}{dT}, \label{eq:thermo:a}\\
    n_i &= \frac{dP}{d\mu_i}, \label{eq:thermo:b}\\
 \varepsilon &= T s  - P + \sum_i\mu_i n_i ~, \label{eq:thermo:c}
\end{flalign}
\end{subequations}
where the temperature $T$ and the set of chemical potentials $\{\mu_i\}$ are the free parameters characterizing the thermodynamic state of the system.
The thermal equilibrium allows us to write any summable thermodynamic quantity marked with $X$ as :
\begin{equation}\label{eq:ThermoSum}
    X_{\rm cosmic}^{(n_f)} = X^{(n_f)}_{\rm QGP} + X_\gamma + X_\nu + X_{e,\mu,\tau} + X_{\rm bosons}
\end{equation}
where the subscripts and their corresponding particle species are self-explanatory. The QCD sector encapsulates deconfined quark and gluon interactions. $n_f$ denotes the number of quark flavors considered; our base data from the microscopical model contains $u, d, s $ quarks. Conveniently, at the QCD transition heavier quarks have almost fully vanished. Recent studies show that heavy charmed hadrons/mesons are formed at the QCD transition \cite{Sharma:2025zhe, Kaczmarek:2025dqt}. However, since the charm quark abundance is already negligible at this temperature scale, the added contribution to the total thermodynamics is negligible. For consistency we still added the data found in \cite{Kaczmarek:2025dqt} to the mix, without any significant effect as expected. Every particle mass was set according to \cite{ParticleDataGroup:2022pth}.
At higher temperature regimes, the inclusion of heavy quarks is required for a realistic picture. It is tempting to straightforwardly add a gas of free quarks to the mix with the argument that the QCD sector at these temperatures is already very close to an ideal gas of free quarks; however, such an approach cannot account for the interaction between the heavy quarks and the gluons. At tree level, we can evaluate the (2+1+1) configuration with (2+1) as follows (from \cite{Borsanyi:2016ksw} Supplementary Information):
\begin{equation}\label{eq:charmCorrection}
    \frac{P^{(2+1+1)}(T)}{P^{(2+1)}(T)}= \frac{\sigma_{\rm SB}(3)+P_{\rm charm}(T)}{\sigma_{\rm SB}(3)},
\end{equation}
where $\sigma_{\rm SB}(n_f)$ is the Stefan-Boltzmann limit for $n_f$ quark flavors considered and $P_{\rm charm}(T)$ the pressure of a gas of free quarks. Note that the Stefan-Boltzmann limit should not only take into account quarks but also gluons, i.e. a non-interacting gas of massive quarks and eight massless gluons. 
A similar treatment to that shown in Eq.~\eqref{eq:charmCorrection} can be applied to extend the description of the QCD sector by adding bottom and top quarks (see Appendix \ref{app:BaseThermo}). 

Between the start and decay of the QGP phase lies another transition: at $T_{EW}\approx 100-200 \rm GeV$ the electromagnetic and the weak interactions split, referred to as the Electroweak phase transition (EWPT). The Higgs vacuum expectation value $v$ is also set at this time; the observed mass of the Higgs boson  $m_H = 125 \rm GeV$ is thought to be high enough to produce a smooth crossover rather than a proper phase transition per se \cite{Kajantie:1996mn,Kajantie:1996qd}. In other words, $v$ evolves continuously from the symmetric phase $v=0$ to the broken symmetry phase with $v=246 \rm~GeV$. In the symmetric phase the cosmic soup is effectively a mix of massless particles. Such a smooth crossover prevents the formation of cosmological relics like domain walls or cosmic strings, which can leave distinct imprints in the SGWB \cite{Correia:2025qif}, see \cite{Hindmarsh:2020hop} for a review.

\begin{figure}
    \centering
    \includegraphics[width=0.8\linewidth]{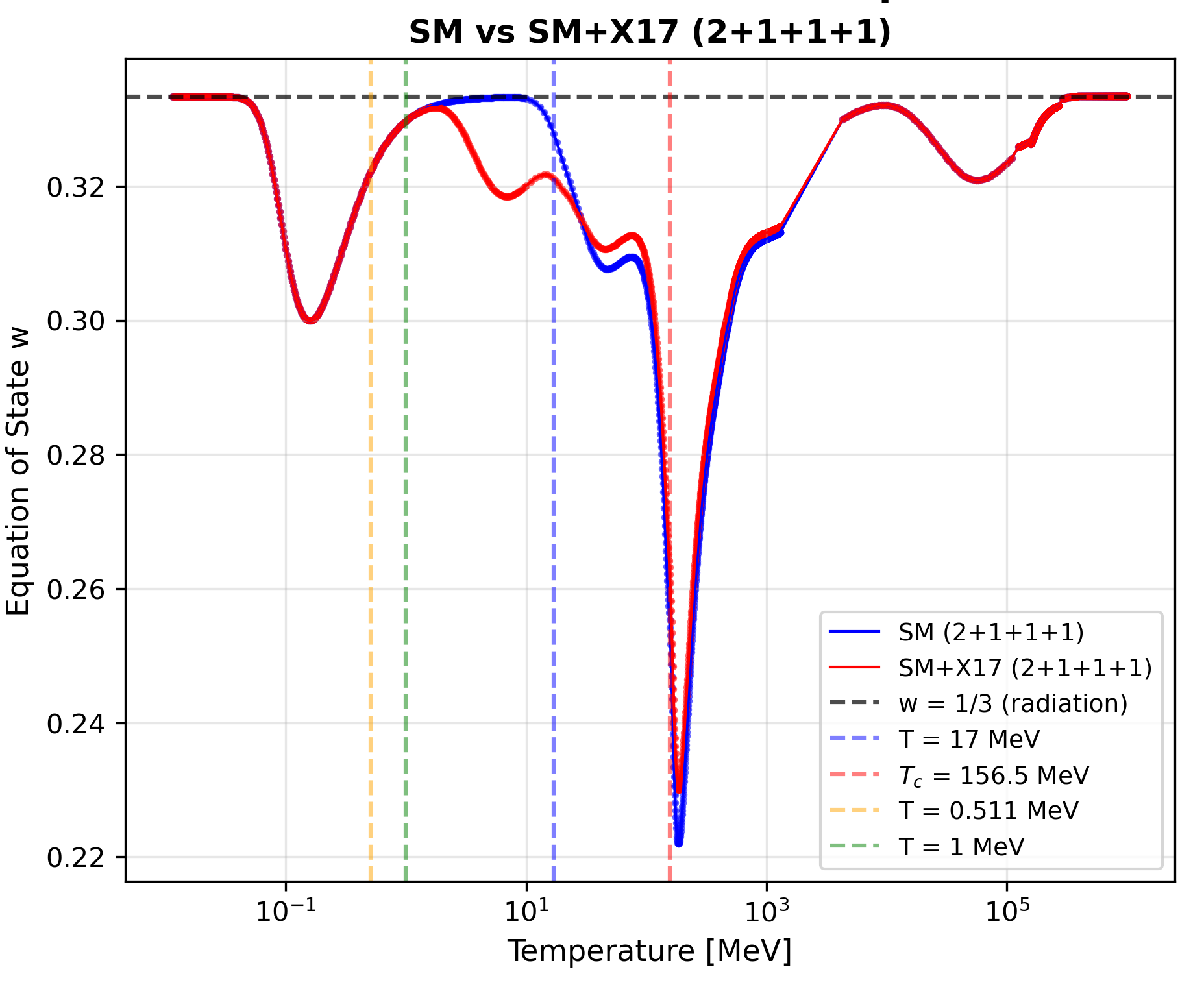}
    \caption{Cosmic EoS tree-level corrected up to the bottom quark (Eq.~\ref{eq:charmCorrection}) for the SM (blue) and SM+X17 case (red). For $T>1300\rm~MeV$, at the EWPT scale we used data from Laine \& Meyer \cite{Laine:2015kra}. For $T<10 \rm~MeV$ we used the \texttt{NUDEC BSM} code \cite{EscuderoAbenza:2020cmq,Escudero:2025kej} to model $\nu$ decoupling. Dashed vertical lines denote key temperatures.}
    \label{fig:EoS_std}
\end{figure}

Phase transitions, whether 1st or 2nd order or a crossover, leave distinct imprints in the EoS. A deviation from the pure radiation value $w=P/\rho=1/3$ occurs there, see Fig.~\ref{fig:EoS_std}.
Here we corrected the thermodynamics up to the bottom quarks. Given the mass of the top quark and the fact that quarks heavier than the strange do not significantly contribute to the QCD transition, ignoring their contribution in the microscopical temperature range $T\in [1:1300]\rm~MeV$ is realistic. 
We see that the QCD transition is the most prominent feature of the EoS in Fig.~\ref{fig:EoS_std}. At the moment corresponding to the confinement of quarks into hadrons and mesons, the entropy and energy density drop, as well as the pressure, though less steeply, creating a dip in the EoS parameter $w$. The same parameter rises to the radiation value $1/3$ after the transition as the leptons and photons become the dominant source of pressure. Then another dip appears at $T\approx 50 \rm~MeV$, corresponding to the annihilation of the newly formed pions. 

Once the QGP has vanished, the other species contributions start to be significant, see Fig.~\ref{fig:partial_component_std}. Neutrinos and photons are added as a relativistic ideal gas of massless fermions and bosons. For charged leptons we use a non-degenerate ideal gas of massive particles approximating the Fermi-Dirac integral with Bessel functions, see Appendix of \cite{Gonin:2025uvc} for details.
The bump in lepton contributions just after the QCD transition in Fig.~\ref{fig:partial_component_std} is due to the muons which have not yet annihilated. Their mass of $m_\mu=105 \rm~MeV$ is close to the critical QCD temperature $T_c$, so that when the QGP has almost fully decayed the charged lepton gas is the main pressure contribution for a short time. Then the muons decay and the charged lepton pressure drops, carried from this point on only by electrons.

At $T\approx 10 \rm~MeV$ the neutrinos start oscillating, the individual lepton flavor asymmetries mix with each other according to the mixing angle \cite{Barenboim:2016shh}, and are fully decoupled at $T\approx 1\rm~MeV$. 
Eventually the electron contribution starts to vanish as electrons annihilate with the positrons at $T\approx 0.511\rm~MeV$ heating up the cosmic components still coupled with photons, mainly hadrons.
This succession of processes also leaves a signature in the EoS parameter, modeled here using the publicly available code \texttt{NUDEC BSM} \cite{EscuderoAbenza:2020cmq, Escudero:2025kej}.

\subsection{Beyond the Standard Model}
\label{subsec:BSM}

In the previous section we described the Universe in its most standard way. The picture is consistent with (most) particle physics observations but does not address some fundamental cosmological conundrums. Unknown or overlooked BSM physics could be the culprit. Let us start from the most grounded observations by mentioning "anomalies" in the SM. Experimental research on BSM physics is a particularly active field; so far, many "anomalies" (most often in the form of resonances) have been found in nuclear and particle physics experiments. The extension in particle physics can occur in the sectors of fermions (spin 1/2), scalar bosons (spin 0), and vector bosons (spin 1). For a recent overview of anomalies in particle physics see Crivellin \& Mellado \cite{Crivellin:2025txc}. In a previous study we looked at the impact of the putative X17 boson \cite{Krasznahorky:2024adr, Alves:2023ree} on the cosmic EoS and the PBH distribution around the QCD transition, see \cite{Gonin:2025uvc}. In Fig.~\ref{fig:EoS_std} we updated the previous study by including $e^+ e^-$ annihilation and the standard crossover at the EW scale. This picture remains close to the standard case, as we do not include new physics per se, but simply add another non-interacting boson to the thermodynamics. It acts as a proof of principle of how particle physics "anomalies" can leave a cosmological imprint.

Turning now to cosmic mysteries, we already mentioned the need for a baryogenesis mechanism able to match the observed baryon asymmetry $b=n_B/s = 8.7 \times 10^{-11}$ inferred from Planck \cite{Planck:2018vyg}. In the SM, $b$ is supposedly linked to the lepton asymmetry $l$ through the Sphaleron process, allowing violation of baryon $B$ and lepton $L$ numbers while conserving $B-L$.
We are not concerned here with the underlying mechanism leading to baryo- or leptogenesis; we refer the interested reader to the reviews and references therein \cite{Bodeker:2020ghk, 2025arXiv250809989V}. Nevertheless, we want to examine the thermodynamic impact of an asymmetric Universe in baryon and lepton number, i.e., accounting for chemical potentials $\mu_B, \mu_Q, \mu_{L_e}, \mu_{L_\mu}, \mu_{L_\tau}$ — respectively the baryonic, electric charge, electronic, muonic, and tauonic sectors. Efficient Sphaleron processes would yield $l=-51/28 \times b$ \cite{Harvey:1990qw}. Such a scenario would remain very close to the picture described in Sec.~\ref{subsec:SM} and the blue curve of Fig.~\ref{fig:EoS_std}, as the asymmetries $b$ and $l$ are very small and the induced chemical potentials do not significantly impact the thermodynamics. It was pointed out in Ref.~\cite{Schwarz:2009ii} that an accurate description of the early Universe requires accounting for those asymmetries, as their introduction along with conservation of the associated charges (see Eq.~\ref{eq:conservation_equations} below) generates \textit{cosmic trajectories} in the QCD phase diagram, see Fig.~\ref{fig:Cosmic_paths}.

\begin{subequations}
\label{eq:conservation_equations}
\begin{align}
    l_\alpha s &= n_\alpha + n_{\nu_\alpha} = n_{L_\alpha}, \label{eq:conservation_equations:a}\\
    bs &= \sum_i B_i n_i = n_B, \label{eq:conservation_equations:b}\\
    qs &= \sum_i Q_i n_i = n_Q \label{eq:conservation_equations:c}
\end{align}
\end{subequations}
where $n_B$ is the net baryon number density, $n_Q$ the net charge number density, $n_{L\alpha}$ the net lepton number, $l_\alpha$ is the lepton flavor asymmetry for a flavor $\alpha=e, \mu,\tau$, and $s$ the entropy density, see equation \ref{eq:ThermoSum}.

Many ideas and BSM processes have been proposed to generate a baryon asymmetry, often able to generate values above observations.
For instance, the Affleck-Dine (AD) mechanism assumes the possibility of inhomogeneous baryogenesis, i.e. there could be patches of the Universe with high baryon number \cite{Allahverdi:2012ju, Kasai:2022vhq}. The AD mechanism also predicts the possibility of a globally high baryon number, which could trigger a 1st order QCD transition. Understanding how the PT occurs in these regimes is necessary, as unexpected processes could evade the constraints on the baryon asymmetry. Boeckel \& Schaffner-Bielich suggested a "little inflation" era (or supercooling) at the QCD crossing for $\mu_B/T>1$ followed by reheating \cite{Boeckel:2010bey, Boeckel:2011yj}. We want to study how these models could evolve thermodynamically, and to do so we use QCD thermodynamics with $\mu_B/T=1,2$ from \cite{Blaschke:2023pqd}.

Another pathway for an early Universe with high baryon chemical potential, which does not require a large baryon asymmetry, is through lepton flavor asymmetries, which recent analyses find to be up to $\mathcal{O}(10^{-2})$ \cite{Oldengott:2017tzj, Kawasaki:2022hvx, Escudero:2022okz, Lattanzi:2024hnq, Li:2024gzf}. 
The idea is straightforward: large values of lepton asymmetries before BBN are hard to constrain, as neutrino oscillations starting at $T=10\rm~MeV$ can change the lepton flavor asymmetries to values fulfilling the constraints \cite{Barenboim:2016shh}. Some values of lepton flavor asymmetries are preferred by CMB and BBN \cite{Froustey:2024mgf,Domcke:2025lzg}. It should be noted that in the references above only neutrino chemical potentials are taken into account in the calculations. The present calculations show non-vanishing chemical potentials for the charged leptons; their contributions are subdominant at $T<10 \rm ~MeV$ which makes the CMB and BBN constraints still relevant, even though not exact. 
Only a direct measurement of the infamously hard-to-measure Cosmic Neutrino Background could shed light on the MeV era.
None of the models presented in this study are favored by neutrino oscillations; the aim of the present paper is to examine how LAU impact the thermal history and link it to PBH formation in the manner of Bödeker et al. \cite{Bodeker:2020stj}. An exhaustive study of the preferred LAU configuration is planned. 

\section{Methods}
\label{sec:method}
After presenting the thermal history after the Big Bang in various contexts, we now turn to how to solve the conservation equations.
In order to solve Eq.~\ref{eq:conservation_equations}, the dependence of the entropy density on the baryonic and electric charge chemical potentials must be evaluated. To do so, following the procedure introduced in Ref.~\cite{Schwarz:2009ii,Wygas:2018otj,Wygas:2019tsx,Formaggio:2025nde}, we rely on a Taylor expansion over the tree-level correction up to the charm quark. We used state-of-the-art QCD pressure derivatives, called susceptibilities, up to 4th order from \cite{Abuali:2025tbd, Kaczmarek:2025dqt}. The lepton sector was solved using the Fermi-Dirac integrals with the JEL approximation \cite{Johns:1996ht}.
Even though our approach is very similar to the latest study \cite{Formaggio:2025nde} and we used the same susceptibilities for the Taylor expansion, our base entropy density $s^{QCD}(T, \mu=0)$ does not come from lattice QCD calculations but rather from the microscopical model \cite{Blaschke:2023pqd}. This comes with the advantage of continuous trajectories, rather than 3 distinct regimes. We also extended the calculations to higher temperatures using polynomial extrapolations from Bresciani et al. \cite{Bresciani:2025vxw} and interpolation from $T=1300 \rm~MeV$ to $T=2000 \rm~MeV$. This interpolation range is arbitrary as the Bresciani and microscopical model temperature ranges overlap; however, a simple concatenation at $T=1300 \rm~MeV$ exhibits a discontinuity.

\begin{figure}
    \centering
    \includegraphics[width=0.6\linewidth]{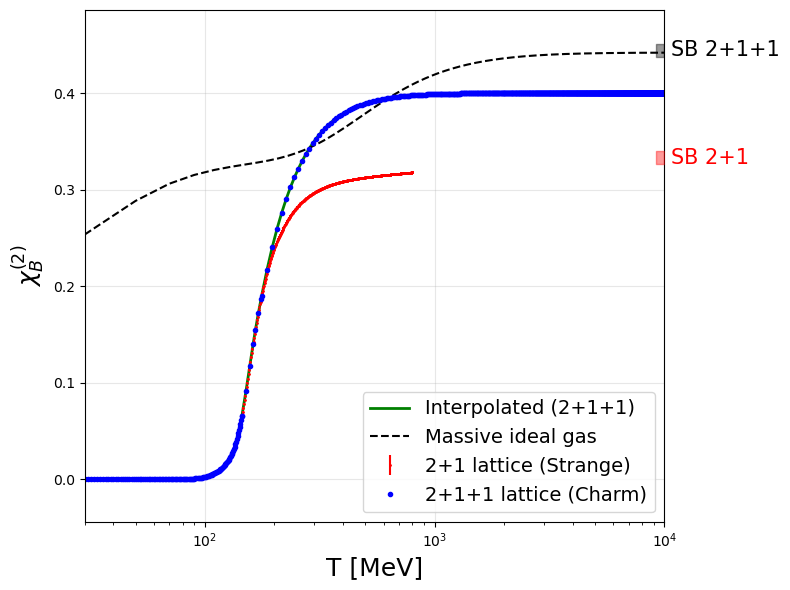}
    \caption{$\chi^B_2$ comparison between the ideal QCD sector made of massive quarks up to the charm and gluons as the black dashed line, the data from Ref.~\cite{Abuali:2025tbd} in red and the derived inclusion of the charm quark using Ref.~\cite{Kaczmarek:2025dqt} as blue dots. The SB limits are shown as colored squares to the right of the plot.}
    \label{fig:chi_B2}
\end{figure}

The (2+1) susceptibilities from Ref.~\cite{Abuali:2025tbd} range from $T\in [30:800]\rm~MeV$; we used the \texttt{Thermal-Fist} package \cite{Vovchenko:2019pjl} for computations below $30 \rm~MeV$.
In order to properly include the charm quark in the calculation we also need the (2+1+1) susceptibilities. Using the susceptibilities from \cite{Kaczmarek:2025dqt} one can derive the relevant susceptibilities$\chi^B_2,\chi^Q_2,\chi^{BQ}_{11}$ $, \chi^B_4, \chi^Q_4, \chi^{BQ}_{13}, \chi^{BQ}_{31}, \chi^{BQ}_{22}$ including the charm quarks. 
We also extrapolate the susceptibilities from Refs.~\cite{Abuali:2025tbd,Kaczmarek:2025dqt}. It can be seen in the figures of the mentioned references that the susceptibilities tend to their Stefan-Boltzmann (SB) limit at asymptotic temperature. We added a point at 90\% of their SB limit at $T=10 \rm ~GeV$ and performed a polynomial fit to ultimately extend the Taylor expansion to higher regimes. We show $\chi^{B}_2$ as an example in Fig.~\ref{fig:chi_B2}. We observe a similar behavior for the (2+1) and (2+1+1) cases, the former being a pure lattice QCD calculation is reliable; the similarities between the red and blue curves show that the inclusion of the charm quark and the polynomial extrapolation are fairly reliable. The comparison with the massive quarks and massless gluons gas shows how poorly this approximation describes the QCD transition.
We stress that the absence of susceptibility data including heavy quarks at high temperature is one caveat of our approach. The present study should motivate the exploration of such regimes.

\section{Results and discussion}
\label{sec:results}
\subsection{Equation of state with LAU and BAU}
In the previous section we presented the equations to solve (Sec.~\ref{sec:EoS}) and how we extended the analysis (Sec.~\ref{sec:method}). We now look in detail at the resulting cosmic trajectories and EoS.

\begin{figure}
    \centering
    \includegraphics[width=0.9\linewidth]{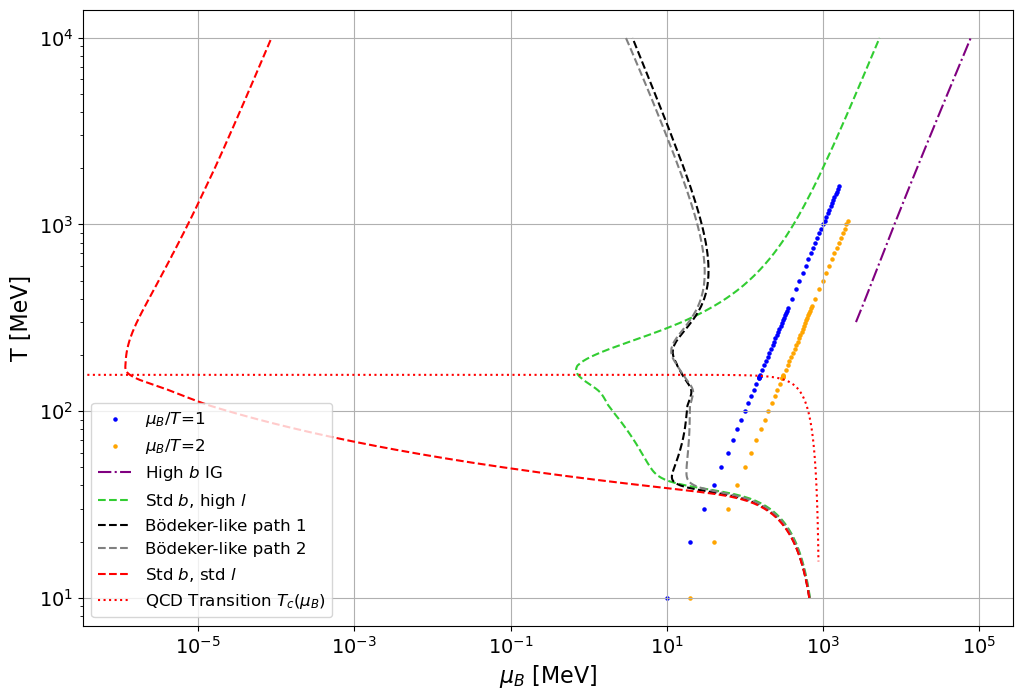}
    \caption{Cosmic trajectories for different scenarios: the dashed lines correspond to realistic scenarios. Colored dots show the trajectories with $\mu_B/T$ imposed. The purple dash-dotted line marked as 'High $b$ IG' corresponds to an ideal gas (IG) calculation with $b=0.1, l_e=-0.1, l_\mu=-l_\tau=0.1$; the lime green dashed line marked as 'Std $b$, high $l$' corresponds to $b=8.6\times10^{-11}, l_e=-l_\mu=l_\tau=-0.1$; the black dashed line marked as 'B\"odeker-like 1': $b=8.6\times10^{-11}, l_e=0, l_\mu=-l_\tau=-0.04$; the grey dashed line marked as 'B\"odeker-like 2': $b=8.6\times10^{-11}, l_e=-0.08, l_\mu =l_\tau =0.04$; the red dashed line marked as 'Std $b$, std $l$': $b=8.6\times10^{-11}, l_e=l_\mu=l_\tau=5.3\times 10^{-11}$. See Appendix for the cosmic trajectories in the $T$ vs $\mu_Q, \mu_{L_e}, \mu_{L_\mu},\mu_{L_\tau}$ planes. The parametrization of $T_c(\mu_B)$ (red dotted) is taken from \cite{Blaschke:2024jqd}.}
    \label{fig:Cosmic_paths}
\end{figure}

The LAU trigger an increase in the absolute value of the lepton chemical potentials. Then electric charge neutrality, Eq.~\ref{eq:conservation_equations:c}, imposes a correlation between the QCD net electric charge $n_Q^{QCD}$ and the lepton sector net electric charge $n_Q^{l}=n_e+n_\mu+n_\tau$.
It follows that the chemical potentials associated with the quarks also increase in absolute value. We can see how the LAU change the cosmic trajectories in the nuclear matter phase diagram in Fig.~\ref{fig:Cosmic_paths}. The dashed grey and black lines correspond to models presented in B\"odeker et al. \cite{Bodeker:2020stj}. The red dashed line, also presented in the same study, corresponds to the case with lepton asymmetry associated with the Sphaleron process $l_\alpha = -51/28\times b$, which we call the standard case 'Std $b$, std $l$'.The dash-dotted line corresponds to an extreme scenario where $b=-l=0.1$, where the Taylor expansion is no longer reliable. For this dash-dotted model we only present the massive non-interacting quark and gluon gas approximation. For easier comparison with the other models, the charm quark is the most massive quark of the model. The quarks and leptons have associated chemical potentials:

\begin{subequations}
\label{eq:part_chem_pot}
\begin{align}
    \mu_{up-type} &= \frac{1}{3}\mu_B + \frac{2}{3} \mu_Q \label{eq:part_chem_pot:a}\\
    \mu_{down-type} &= \frac{1}{3}\mu_B - \frac{1}{3} \mu_Q \label{eq:part_chem_pot:b}\\
    \mu_{\nu_\alpha} = \mu_{L_\alpha} &= \mu_{\alpha^{\pm}} + \mu_Q \label{eq:part_chem_pot:c} 
\end{align}
\end{subequations}

where $\mu_Q$ is the electric charge chemical potential, $\mu_{\nu_\alpha}$ (also written $\mu_{L_\alpha}$) corresponds to the neutrino flavor $\alpha$ chemical potential and $\mu_{\alpha^{\pm}}$ the charged leptons.
The gluons are treated without chemical potential as an ultra-relativistic gas of massless bosons.
All the dashed lines in the figures throughout this study are determined for $b=8.6\times10^{-11}$ from \cite{Planck:2018vyg}, $q=0$ consistent with constraints on the electric charge of the Universe \cite{Caprini:2003gz}, and $l_\alpha$ arbitrarily fixed, corresponding to realistic trajectories found by solving the conservation equations \ref{eq:conservation_equations}.

Since the Taylor expansion breaks down when $\mu_B/T \approx 0.1$, the extreme dashed-dotted model is computed only at high temperatures using the ideal gas approximation. Furthermore, a pion condensate can form when $\mu_Q > m_\pi$ or $|l_e+l_\mu|\gtrsim 0.1$, which could leave an imprint in the SGWB \cite{Middeldorf-Wygas:2020glx, Vovchenko:2020crk, Ferreira:2025zeu, 2025arXiv251111995D}. The entry into and departure from the condensate could be a 1st or 2nd order PT depending on the value of $|l_e+l_\mu|$. We show explicitly the pion mass in the T vs $\mu_Q$ plane in the Appendix Fig.~\ref{fig:Tvsmu_Q}. None of the realistic trajectories presented form a pion condensate. The $\mu_B/T$ fixed cases also do not form a pion condensate because $\mu_Q$ is null in the microscopical model.

Gao \& Oldengott \cite{Gao:2021nwz} used a functional QCD method to explore cosmic trajectories with high lepton flavor asymmetry. They find the signature of a 1st order QCD transition when looking at the quark chemical potential \cite{Gao:2021nwz, Lu:2023msn, Gao:2024fhm}. The condition for a 1st order PT relies on the determination of the Critical End Point (CEP), where the PT stops being a crossover. They find that if $|\mu_q|>200 \rm~MeV$ at $T_{CEP}=118 \rm~MeV$ the PT is 1st order. None of the realistic cosmic trajectories fulfill this condition.

While it is clear that $\mu_B/T$ fixed paths are not realistic around the QCD transition, they remain interesting when considering AD baryogenesis or high lepton flavor asymmetries. Such trajectories could approximate the thermodynamic evolution of the models proposed by Boeckel \& Schaffner-Bielich \cite{Boeckel:2010bey, Boeckel:2011yj} or the High Baryon Bubbles proposed in \cite{Kasai:2022vhq}. These regions could retain a high baryon number and somehow evolve independently if they are large enough, i.e. on superhorizon scales. We want to keep the number of assumptions as minimal as possible, so we do not include lepton asymmetry in the calculations of these models; we take the data from figures 8 and 12 of \cite{Blaschke:2023pqd} and insert it in Eq.~\ref{eq:ThermoSum}. The paths with $\mu_B/T$ fixed use (2+1) quark flavors while the dashed lines use tree-level correction to account for the charm quark too. The reason lies in Eq.~\ref{eq:charmCorrection}: the accurate determination of $P_{charm}$ in the presence of chemical potentials would have to take into account the electric charge chemical potential, see Eq.~\ref{eq:part_chem_pot}. Since $\mu_Q$ depends on the lepton asymmetry and we chose to use a simplistic model, we shall remain conservative and not add any tree-level correction. The charm quark is not thought to be very significant at the QCD transition \cite{Gonin:2025uvc, Kaczmarek:2025dqt}; the leptons, however, are, and with $\mu_B/T$ fixed we do not compute the lepton chemical potentials — we discuss the effect of this choice in the following paragraphs.
It is worth noting that at $T\gtrsim500 \rm~MeV$ the slopes of the $\mu_B/T$ fixed paths are similar to those of the 'High $b$ IG' and 'Std $B$, high $l$' models. This is because with massless particles, which is a reasonable approximation away from the QCD transition, the charge conservation equations \ref{eq:conservation_equations:b} and \ref{eq:conservation_equations:c} are analogous to $\mu_B/T = const$. In the appendix of \cite{Formaggio:2025nde} they find $\mu_B/T \approx 65b - 12.5l$ valid for $l\gg b$.

\begin{figure}
    \centering
    \includegraphics[width=0.9\linewidth]{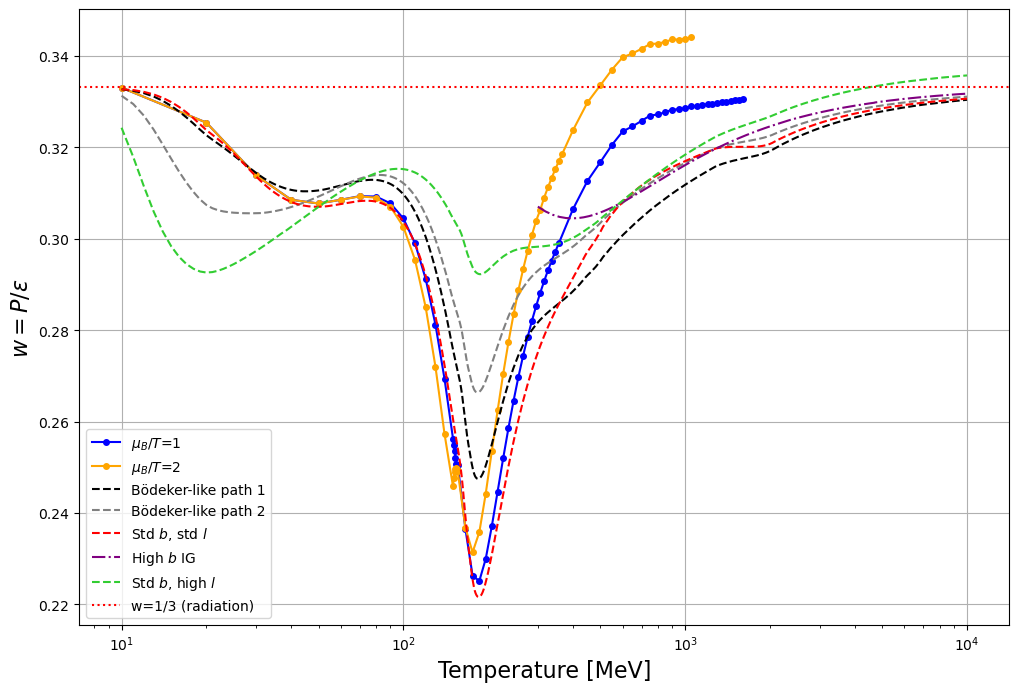}
    \caption{EoS corresponding to the different cosmic trajectories. The same color code from Fig.~\ref{fig:Cosmic_paths} is applied. Flat portion in the dashed lines around $T\approx 1-2GeV$ is only due to interpolations of our data to the polynomial extrapolation from \cite{Bresciani:2025vxw}. The little zig-zag in $\mu_B/T=2$ just after the QCD transition is an artifact of the present version of the microscopical model \cite{Blaschke:2023pqd}.
    %\textbf{@OLEKSII maybe add something?}
    }
    \label{fig:EoS_muB}
\end{figure}

The EoS of the different models is plotted in Fig.~\ref{fig:EoS_muB}. The dashed grey, black and red lines reproduce the B\"odeker results; the dip in the EoS at the QCD transition is mitigated by the baryon chemical potential but also by the lepton chemical potentials.
It is clear from the colored continuous lines that the baryon chemical potential does provide a mitigation of the softening, but the difference with the 'lepton powered models' shows that the lepton sector might be even more important when it comes to creating a shallower dip at the QCD transition. We explain this through the $\mu$ lepton, in the two B\"odeker-like and 'Std $b$, high $l$', $l_\mu$ is large, hence $\mu_{L_\mu}$ follows, which
makes the contribution of the $\mu$ sector more significant, see Fig.~\ref{fig:partial_component_high_b_high_l}. At $T=156 \rm~MeV$ the $\mu$ have not yet decayed, and can therefore still compensate for the dip created by the QCD transition. 

\begin{figure}
    \centering
    \includegraphics[width=\linewidth]{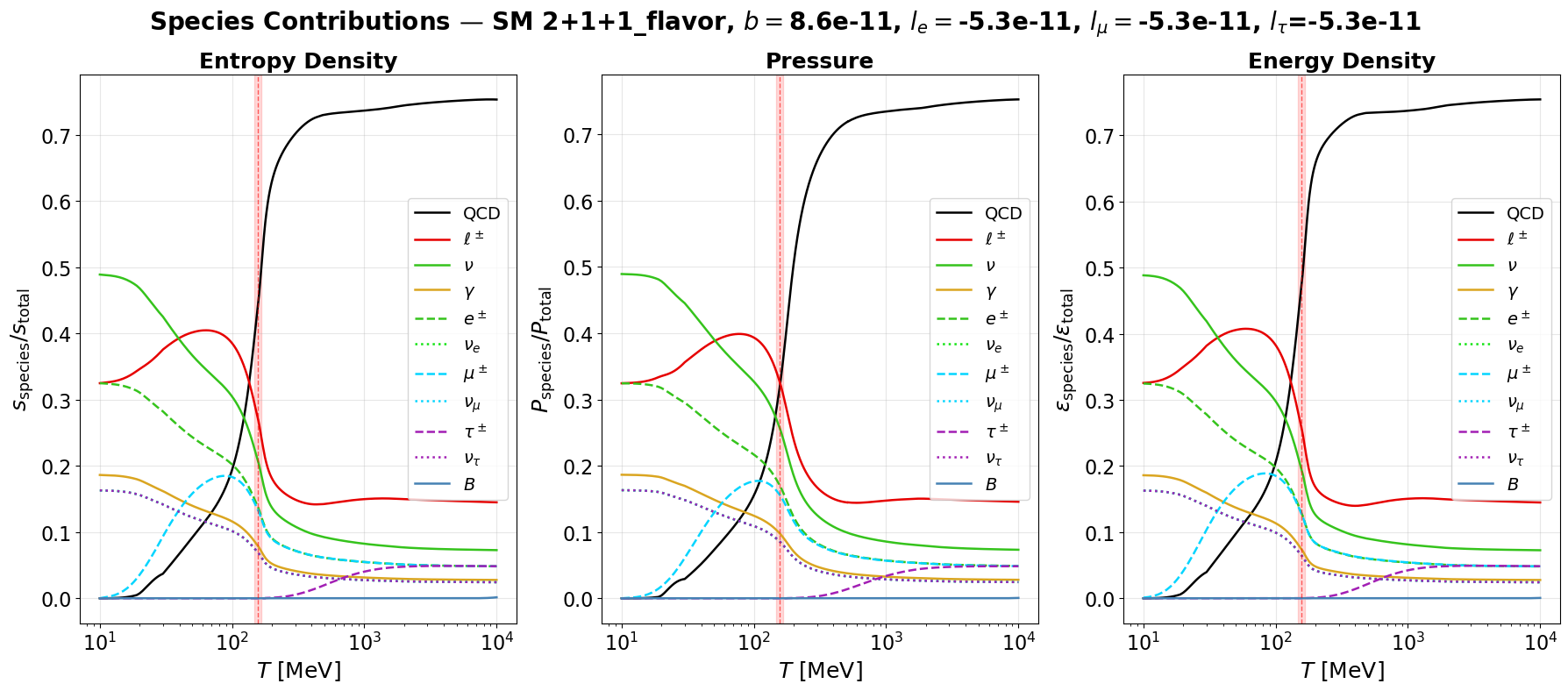}
    \caption{Partial thermodynamic contributions from different particle species and flavors in the standard case 'Std $b$, std $l$' with small asymmetries. The red shaded region highlights the QCD transition pseudo-critical temperature $T_c=156.5$. The different species contributions can be read in the legend; '$B$' denotes bosons.}
    \label{fig:partial_component_std}
\end{figure}

\begin{figure}
    \centering
    \includegraphics[width=\linewidth]{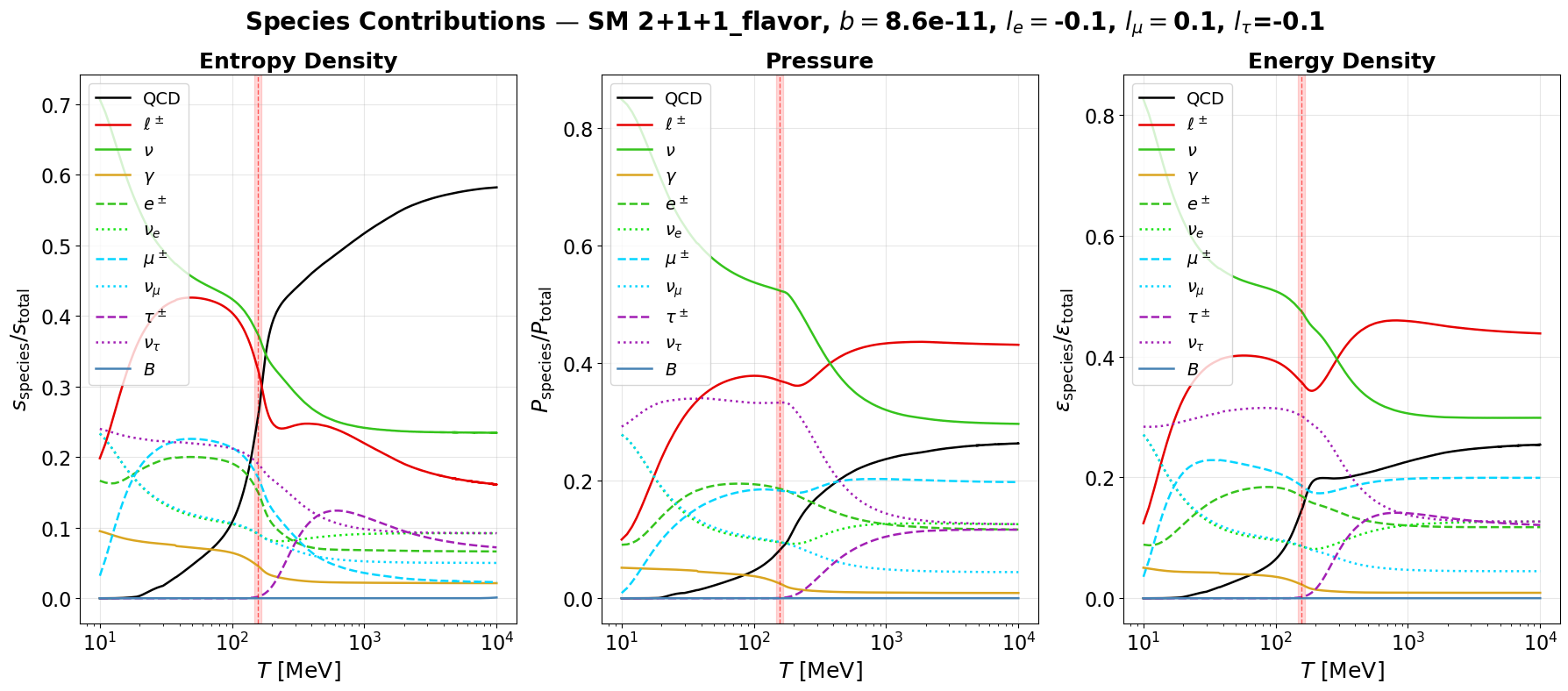}
    \caption{Partial thermodynamic contributions from different particle species and flavors in the case 'Std $b$, high $l$' with high lepton flavor asymmetries.}
    \label{fig:partial_component_high_b_high_l}
\end{figure}

In Fig.~\ref{fig:partial_component_high_b_high_l}, the effect of high $l_\alpha$ on the $\nu_\alpha$ contribution is clear. Looking closely at the purple lines, we see how the decay of the $\tau$ charged leptons is associated with the rise of the $\nu_\tau$ as the '$\tau$ leptonic' charge must be conserved. Comparing with Fig.~\ref{fig:partial_component_std}, we see that the QCD sector does not necessarily dominate the pressure or energy density at high $T$ in the presence of LAU. Furthermore, the behavior of the different species contributions is not identical between different thermodynamic quantities. In 'Std $b$, std $l$' charged leptons dominate shortly after the QCD transition; with the 'Std $b$ high $l$' model, neutrinos are the dominant contributors to the entropy density after the QCD transition. The charged leptons might come close to the neutrino contribution, but the decay of $\mu$, associated with a rise in $\nu_\mu$ contribution, leaves the charged lepton contributions subdominant. For the same model there is a sharp drop in entropy and energy density at the QCD transition, corresponding to the zig-zag in the associated cosmic trajectory Fig.~\ref{fig:Cosmic_paths}.

Increasing $\mu_B/T$ also comes with the effect of shifting the minimum of the softening to lower temperatures; this is not observed in the presence of lepton asymmetry because $\mu_B/T_c$ is small. The effect is not very significant, but it can be explained by the well-known shift of $T_c$ to lower value as $\mu_B$ increases \cite{HotQCD:2018pds,Blaschke:2024jqd}, shown in Fig.~\ref{fig:Cosmic_paths} as the red dotted line.
$\mu_B/T=1$is close to the standard case, and the minimum is barely distinguishable. The dip to the minimum is steeper compared to the B\"odeker-like and standard models, which can be partially explained by the difference in the number of quark flavors, see figure 2 from \cite{Gonin:2025uvc}. The inclusion of heavier quarks would make the rise to $w=1/3$ slower, because massive quarks such as the bottom are not yet ultra-relativistic.

One more notable feature of the continuous colored lines appears at high temperature: $w$ rises above the radiation value, corresponding to a Universe 'stiffer' than what can be expected with pure radiation. Such an effect is not observed in the B\"odeker-like models, because the baryon chemical potentials are not high enough for departures from pure radiation at high temperatures. The purple dash-dotted line, where the QCD sector is an ideal gas of massive quarks and gluons, also sits around $w=1/3$ at high temperature; even though it has a large chemical potential, the Fermi-Dirac distribution imposes $w=1/3$ in any case. The lime green dashed model also shows $w>1/3$ at high temperature, which can be explained by the fact that this model also exhibits high chemical potential at high temperature, triggering the departure from the ideal gas.

\begin{figure}
    \centering
    \includegraphics[width=0.9\linewidth]{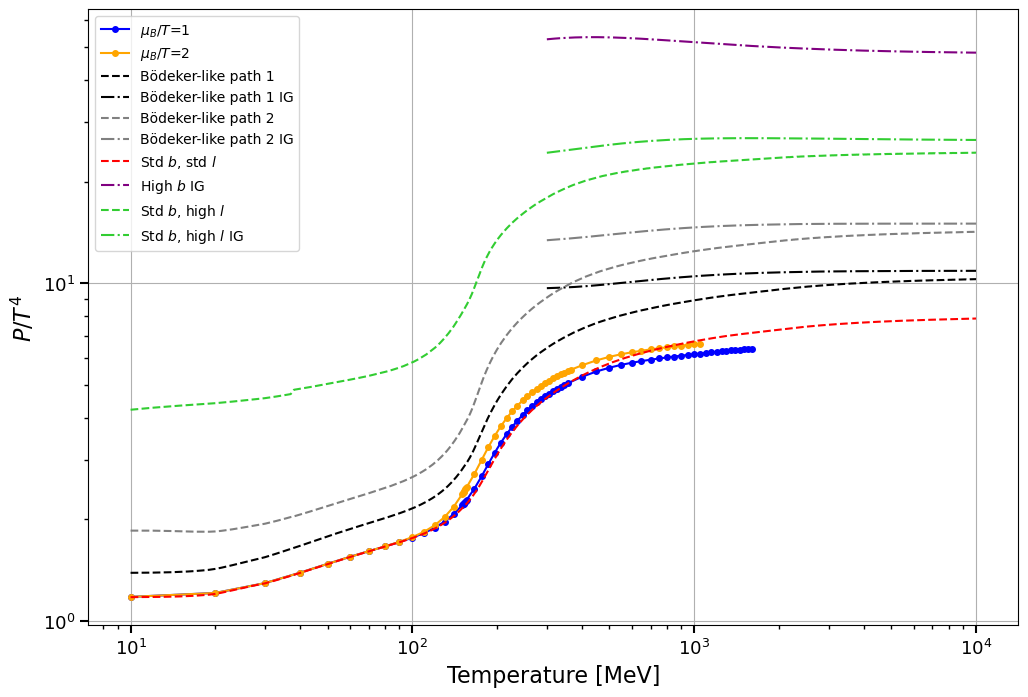}
    \caption{Normalized pressures $P/T^4$, the same color code from Fig.~\ref{fig:Cosmic_paths} is applied. The dash dotted lines correspond to the ideal gas cases. The slight discontinuity in 'Std $b$, high $l$' around $T=30-40 \rm ~MeV$ correspond to the concatenation of the microscopical model with \texttt{Thermal-Fist}.}
    \label{fig:Pressures}
\end{figure}

Since $w$ is simply defined as the ratio of pressure over energy density, $w=P/\varepsilon=1/3$ is not a limit indicating incorrect physics. In Fig.~\ref{fig:Pressures} we plot the pressures of the different models and their associated IG approximations, which act as upper limits. It is clear that all the models sit below the IG limits. This is the reason why we set the susceptibilities to 90\% of their SB limit. Using the exact SB limit would be incorrect, as it would extend the thermodynamics to their asymptotic limit, typically overshooting the value of the Taylor expansion. While there is no physical motivation for taking 90\% of the SB values, such an approach avoids overshoots.
The behavior of the energy density is similar. The purple 'High $b$ IG' shows the highest pressure, as it is also the model with the highest chemical potential. It might be surprising not to find the fixed $\mu_B/T$ mmodels on top of the others, but we are plotting the total pressure from Eq.~\ref{eq:ThermoSum} and these models do not account for lepton chemical potential; hence they might not even exhibit higher pressure values than the 'Std $b$, std $l$' model. 
This shows how important it is to take into account the conservation equations \ref{eq:conservation_equations}.

Having established the cosmic trajectories and their thermodynamic consequences, we now turn to the PBH mass distributions induced by the different LAU.

\subsection{The Primordial Black Hole mass spectrum}

In this section we briefly reintroduce the formalism to translate the EoS into a PBH distribution. For a detailed discussion see \cite{Byrnes:2018clq,Musco:2023dak,Gonin:2025uvc} and the references cited in the following paragraph.
Primordial black holes are a unique probe of the early Universe; since these objects form in the radiation era, they can provide a picture of the Universe before the CMB. Notably, this was proposed as a probe for lepton flavor asymmetry in \cite{Bodeker:2020stj}. We argue here for a similar idea, computing the PBH mass spectrum from the EoS presented in Section \ref{sec:EoS} and comparing the induced distributions and their implications for future observations.
Fluctuations of energy density in the very early Universe are necessary to explain the current energy distribution of the Universe. Hawking and Carr introduced the idea that some of these early fluctuations could have been large enough to collapse to PBHs when the particle horizon crosses the fluctuation radius at $t_{\rm cross}$ \cite{Carr:1974nx}. Although early energy density fluctuations may have many possible origins, the main scenario is cosmic inflation, which typically assumes a Gaussian fluctuation spectrum.
No matter what the source of the fluctuations is, they should be larger than the Jeans length at maximum expansion if they are to undergo gravitational collapse. Recently, Carr explored different formation scenarios \cite{Carr:2020xqk,Carr:2020gox}, and various studies provide numerical simulations of the collapse of overdensities \cite{Escriva:2022bwe,Musco:2023dak,Escriva:2019nsa,Musco:2004ak,Musco:2012au}. We refer interested readers to the 2021 review by Escrivà \cite{Escriva:2021aeh} and the references therein. These studies find that a PBH does not form exactly at the horizon mass, which has the notable effect of widening peaks in the distributions; the QCD peak for instance sees its width increase. For simplicity, we stay in the purely analytical formulation as we are interested in drawing a general picture of the PBH spectrum. We assume that a PBH forms exactly at the horizon mass $M_H = M_{PBH}$.
Collapse can occur at different epochs during the radiation-dominated era, depending on the size of the fluctuation. Following Eq.~4 in B\"odeker et al. \cite{Bodeker:2020stj}, the particle horizon grows as follows:
\begin{equation}
\label{eq:horizonMass}
    T \approx 700~g_\varepsilon^{-1/4} \sqrt{M_\odot/M_H}~{\rm MeV}~.
\end{equation}
It is possible to account for the critical behavior of a collapse to a PBH, i.e. taking into account the fact that PBHs are not born with exactly $M_H$, through the introduction of a factor $\gamma$ \cite{Braglia:2021wwa}. As mentioned above, for simplicity we set $\gamma=1$. A different value of $\gamma$ would slightly shift the peaks to higher or lower masses. We plan a more realistic study using a simulation-based PBH spectrum.
The size of candidate PBH fluctuations increases with time; a full PBH mass spectrum is possible from Planck's mass ($10^{-5}$ g) if formed at Planck's time ($10^{-43}$ s) to supermassive range ($10^5 M_\odot$) for those formed as late as $1$ s after Big Bang \cite{Carr:2020gox}. 
It should be noted that Eq.~\eqref{eq:horizonMass} depends on $g_\varepsilon$, which is a thermodynamic function, so the horizon mass differs slightly between the different models. For a detailed comparison between the SM and SM+X17 scenarios, see Appendix of \cite{Gonin:2025uvc}.
We follow Carr's prescription to fully describe an overdense region by its energy density contrast:

\begin{equation}\label{delta}
    \delta = \frac{\varepsilon-\varepsilon_b}{\varepsilon_b}~,
\end{equation}
where $\varepsilon$ is the energy density in the region and $\varepsilon_b$ is the background energy density. Then the fate of an overdensity is determined by the relation between $\delta$ and the threshold $\delta_c$ \cite{Carr:1975qj}. If $\delta > \delta_c$ at $t_{cross}$ a PBH is formed, if not the overdensity is eventually dispersed away by the pressure.
Assuming Gaussian fluctuations, the fraction of the Universe collapsing is
\begin{equation}\label{eq:beta}
    \beta(M)\approx \mathrm{erfc}\left[\frac{\delta_c(w(T(M))}{\sqrt{2}\delta_{\rm rms}(M)}\right]~,
\end{equation}
where $M$ is the PBH mass, $\rm erfc$ is the complementary error function and $\delta_{\rm rms}$ is the root mean square amplitude of the Gaussian fluctuation. Following \cite{Bodeker:2020stj,Carr:2019kxo},
\begin{equation}\label{eq:delta_rms}
    \delta_{\rm rms} = A\times(M/M_\odot)^{(1-n_s)/4},
\end{equation}
where $n_s=0.97$ is the spectral index taken at its CMB value \cite{Planck:2018vyg}. The amplitude $A$ is a normalization parameter that expresses the strength of the fluctuations.

The present fraction of DM in a PBH of mass $M$ is then
\begin{equation}
\frac{df_{\rm PBH}(M)}{d \ln M} \approx 2.4 ~\beta(M)\sqrt{M_{\rm eq}/M}~,
\end{equation}
where $M_{\rm eq}$ is the horizon mass at matter-radiation equality. The numerical factor originates from
$ 2.4=2(1+\Omega_b/\Omega_{\rm CDM})$, with $\Omega_b=0.0456$ and $\Omega_{\rm CDM}=0.245$ being the baryon and CDM density parameters from \cite{Planck:2018vyg}. 

\begin{equation}\label{eq:f_pbh}
    f_{\rm PBH} \equiv \int_{M_{\rm min}}^{M_{\rm max}} \frac{df_{\rm PBH}}{d \ln M} \,d \ln M\,.
\end{equation}

\begin{figure}
    \centering
    \includegraphics[width=\linewidth]{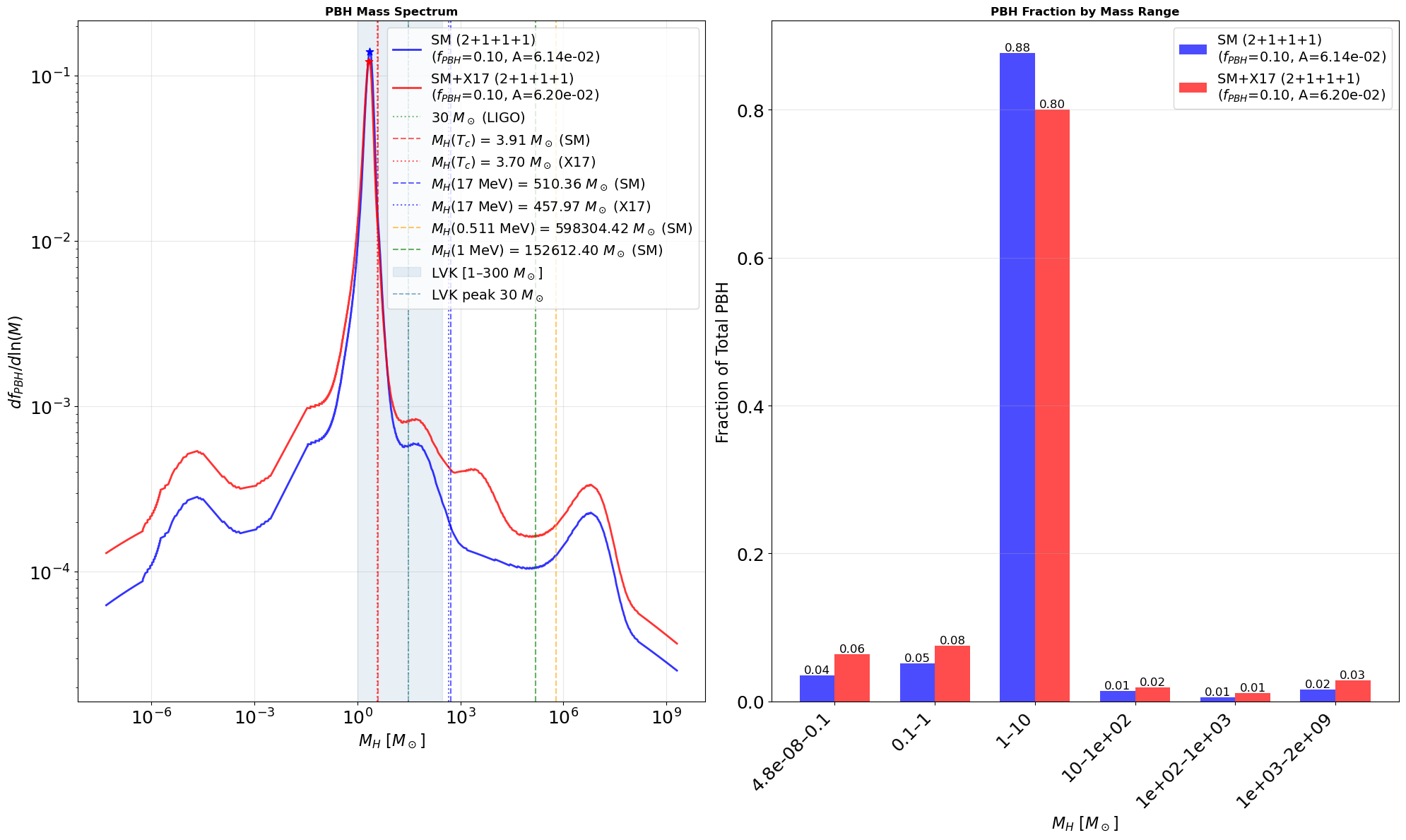}
    \caption{Comparison of SM and SM+X17 PBH mass spectra without chemical potentials. The legend gives $f_{PBH}$, the fraction of DM in PBHs, and $A$, the normalization amplitude of $\delta_{rms}$. The stars denote the maximum of the distribution; the color code of the vertical lines is the same as in the Fig.~\ref{fig:EoS_std}}
    \label{fig:PBH_spectrum_SM_X17}
\end{figure}

The dips in the EoS are of particular interest for PBH formation, as they can be understood as a softening of the Universe: they increase the chance that a given overdense region collapses into a PBH.
In Fig.~\ref{fig:PBH_spectrum_SM_X17} we show the PBH mass spectrum for the SM and SM+X17, including now the super-massive black hole (SMBH) production at $e^+e^-$ annihilation as well as the sub-solar peak at the EWPT. The logarithmic scaling can be deceiving as the SM+X17 seems to produce many more PBHs outside of the QCD peak. The slight mitigation of the QCD softening from X17 does not drastically modify the mass spectrum, but it does show an impact on the entire PBH spectrum.
Although we do not show the PBH spectrum for $\mu_B/T$ fixed, — since the absence of lepton chemical potential makes the behavior of the QCD transition unreliable — we can predict that the QCD peak would be mitigated, with the rest of the distribution increasing its contribution accordingly. The case would not be as straightforward; since $\mu_B/T = 2$ shows a stiff EoS prior to the QCD transition, thus the production in this era would be harder and PBH production mitigated.
The increased production post-QCD transition seen for SM+X17, 'B\"odeker-like' and 'Std $b$, high $l$', could have significant observational effects; we discuss this mass range extensively in \cite{Gonin:2025uvc} Sections 3 and 4.

\begin{figure}
    \centering
    \includegraphics[width=0.9\linewidth]{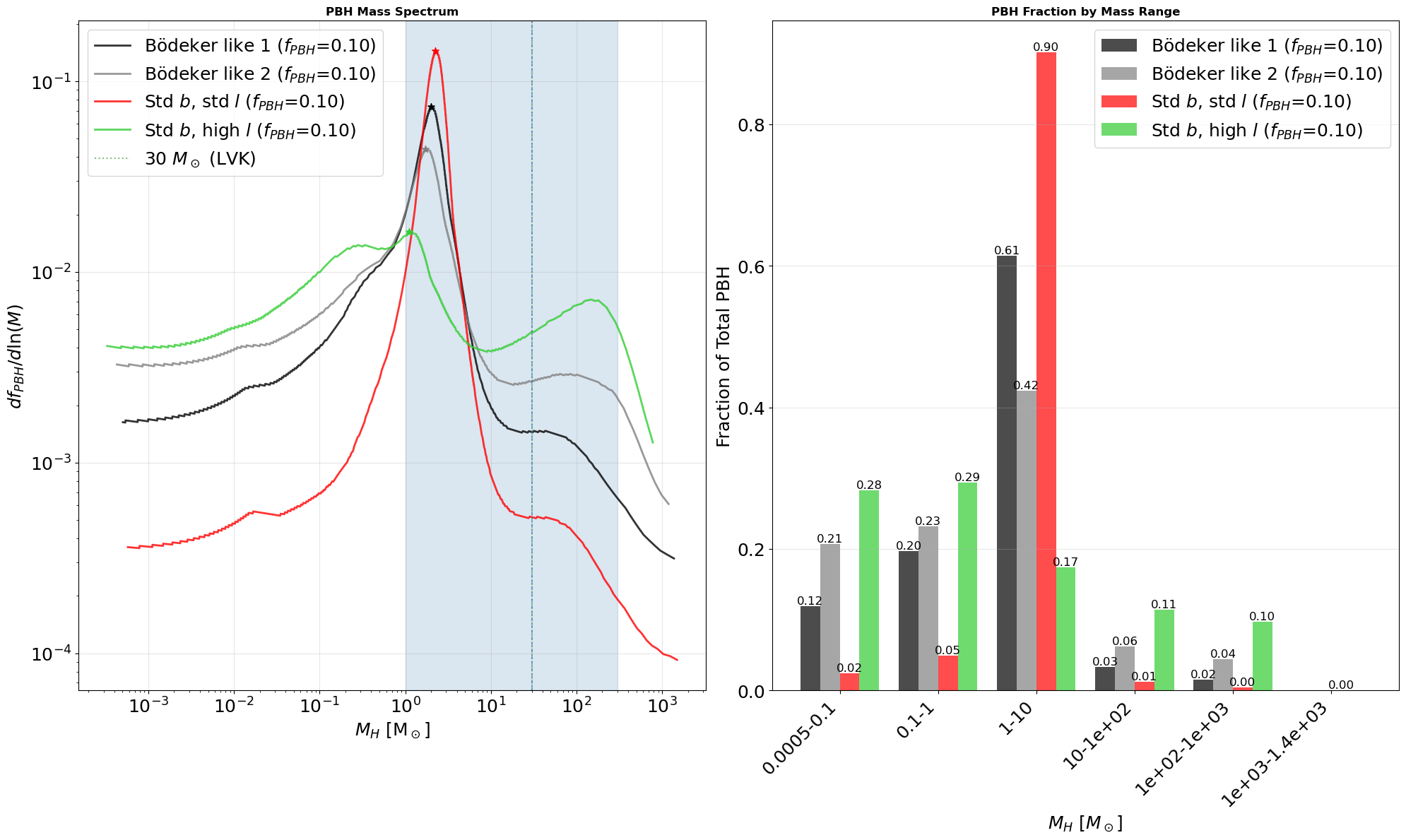}
    \caption{B\"odeker-like PBH spectra and the 'Std $b$, high $l$'. Using the same color code as Figs.~\ref{fig:Cosmic_paths}, \ref{fig:EoS_muB} and \ref{fig:Pressures}. The shaded region shows the observations of GWTC-4 \cite{2025arXiv250818082T}, including the analysis from Ruiz-Rocha et al. \cite{Ruiz-Rocha:2025yno} up to $M_{BH}\sim 300~M_\odot$.}
    \label{fig:pbh_Bodeker-like}
\end{figure}

The standard case, B\"odeker-like models and 'Std $b$, high $l$' are shown in Fig.~\ref{fig:pbh_Bodeker-like}. In the figure 3 of B\"odeker et al. \cite{Bodeker:2020stj} similar spectra are plotted but the normalization of $f_{PBH}$ is not consistent between the models. Here we plot them again with $f_{PBH}=0.1$. Again, the mitigation of the QCD peak triggers increased production in other eras, similarly to the SM+X17 case. 'Std $b$, high $l$' might be the most notable one, the high $\mu_B$ at high $T$, associated with a stiffening of the EoS,  further increases the production of PBHs. The formation of a dome pre-QCD transition at $M_H \sim 2\times10^{-1} M_\odot$ is caused by the decay of the $\tau$ lepton, leaving a deep imprint in the thermal history, see Fig.~\ref{fig:partial_component_high_b_high_l}. 
For models carrying a large tau lepton chemical potential, $\tau^+ \tau^-$ annihilation  helps to soften the EoS pre-QCD transition, but the onset of the $\nu_\tau$ contribution to carry $l_\tau$ mitigates the QCD drop and hence the $M_{PBH}\sim 1~M_\odot$ production. The neutrinos dominate $P$ and $\varepsilon$ after the QCD transition, so the behavior of the EoS and PBH production comes from the leptonic sector. 
At $T\sim 100 ~\rm~MeV$ $\sim m_\mu$ the $\mu^+ \mu^-$ annihilate and completely disappeared by $T\sim 20 ~\rm~MeV$, creating another peak at $M_H\sim 200 M_\odot$. The number density of $\nu_\mu$ increases and the neutrinos become the dominant thermodynamic contribution. 

As mentioned in Section \ref{sec:EoS}, a pion condensate can form in the presence of high lepton asymmetry, with impact on the PBH distribution \cite{Vovchenko:2020crk}. Furthermore, cosmological relics from 1st order transitions could also collapse into PBHs or interact with them \cite{Khlopov:2000js,Baker:2021nyl,Vilenkin:2018zol}. PBHs have also been proposed as a probe for a 1st order EWPT \cite{Hashino:2022tcs}.

\subsection{Discussing the constraints and positive evidence}
After showing how the PBH mass spectrum is modified in the presence of LAU and BSM anomalies like X17, we discuss how these results compare with current observations.
We should start by stating that the aim of the present paper is not to constrain the models presented, nor to make quantitative predictions for future observations. We are mainly interested in describing the EoS behavior around the QCD transition across a diverse variety of models. The PBH mass spectrum is another way to look at the consequences of modifications of the EoS, and would become a direct probe of the Universe pre-BBN should future observations confirm the existence of PBHs.

The stiffening of the EoS pre-QCD transition associated with sub-solar mass PBH formation has implications for microlensing observations. Although it is hard to draw firm conclusions because different collaborations claim different results and constraints, it remains worth mentioning as it is an active field of research. Moreover, the constraints rely on assumptions which are not fully secured. For instance, the spatial distribution of PBHs could have a significant impact on the constraints; see the recent review by Green \cite{2026arXiv260215974G}. Hawkins in Ref.~\cite{Hawkins:2015uja} discussed how the galactic rotation curve can bias microlensing observations.
The current discussion in a nutshell is as follows: in the late 90s and early 2000s, the MACHO collaboration claimed observations compatible with a significant fraction of the Milky Way halo in sub-solar compact objects \cite{MACHO:1996qam,MACHO:2000qbb}. EROS \cite{EROS-2:2006ryy} and OGLE \cite{Mroz:2024wag,Mroz:2025xbl} were not able to corroborate the results. Hawkins \& Garcia-Bellido \cite{Garcia-Bellido:2024yaz,Hawkins:2025mlo} showed how using the Milky Way rotation curve inferred by GAIA can mitigate the constraints from OGLE.
To add to the debate, the Subaru telescope can also be used to search for microlensing events in Andromeda; in their 2017 analysis they found strong constraints \cite{Niikura:2017zjd} on the mass range $M_{PBH}\sim 10^{-11}-10^{-5} M_\odot$.  A 2026 analysis \cite{2026arXiv260205840S} mitigates this original claim, explicitly stating that it originated from a better treatment of the data.
Another way to constrain this mass range is through PBH-star interactions \cite{Esser:2022owk,2026arXiv260223429E}.
For SMBHs the main constraint comes from $\mu$ distortions of the CMB \cite{Nakama:2017xvq}. It is crucial to include the SMBH peak for models with LAU; we leave this to future studies.

\begin{figure}
    \centering
    \includegraphics[width=\linewidth]{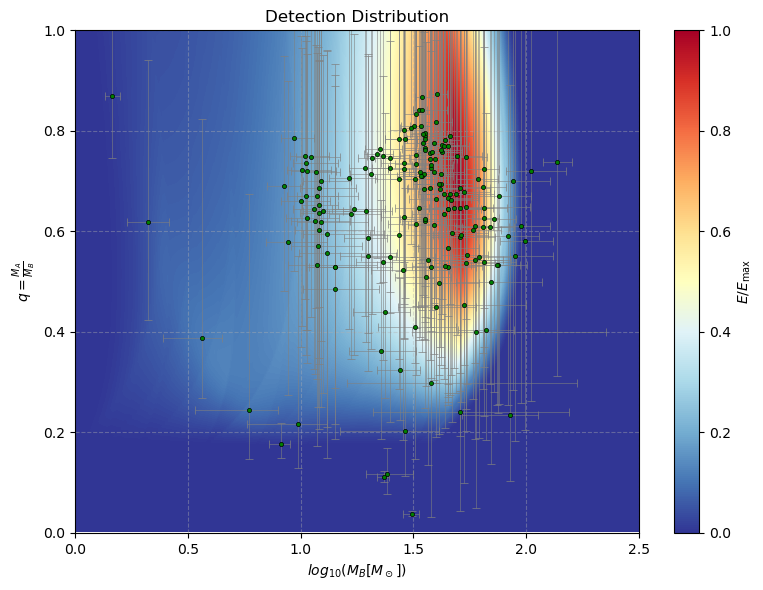}
    \caption{Probability density of PBH mergers with masses $M_A<M_B$. Inspired by Figures 4, 5 and 6 from B\"odeker et al.\cite{Bodeker:2020stj}. Events from GWTC-4 are shown as green dots \cite{2025arXiv250818082T}. We used O4 strain noise from \cite{Essick:2025zed}.}
    \label{fig:pbh_proba_density}
\end{figure}

Stellar mass PBH production has direct implications for GW observations; see Ref.~\cite{LISACosmologyWorkingGroup:2023njw,2025GCN.42724....1S} for a recent review.
As shown by B\"odeker et al. in Ref.~\cite{Bodeker:2020stj} lepton asymmetry can be a convenient way to make GW observations agree with the predicted PBH mass distribution from thermal history. We showed the current LVK observation mass range in Fig.~\ref{fig:PBH_spectrum_SM_X17} and \ref{fig:pbh_Bodeker-like}. In Fig.~\ref{fig:pbh_proba_density}, we show how the probability density of PBH mergers compares to the actual events from GWTC-4 \cite{2025arXiv250818082T}; to model the PBH merger rate we assume only late binaries \cite{Clesse:2016ajp}. Comparing with Figures 4, 5 and 6 from Bödeker et al.~\cite{Bodeker:2020stj}, it is clear how the lepton asymmetry can shift the peak in the probability density of PBH mergers closer to the observed aggregate in the $q-M_B$ plane. Where $q$ is the mass ratio of the binary and $M_B$ the heaviest binary component

The sub-solar mass range is particularly interesting: on November 12, 2025, the LVK collaboration reported a candidate sub-solar mass merger event \cite{LISACosmologyWorkingGroup:2023njw,2025GCN.42724....1S}. It was quickly followed by an analysis finding the merger rate of such an event to be compatible with a slightly modified version of 'B\"odeker-like 2'\footnote{The spectrum they present is tilted compared to the one from this study and Bödeker's, which typically appears when modifying the value of $n_s$ in Eq.~\ref{eq:delta_rms}} \cite{2026arXiv260106024C}. 
PBH formation and the peaks associated with thermal history can leave hints in the SGWB \cite{Braglia:2021wwa,Braglia:2022icu}.

%%%%%%%%%%%%%%%%%%%%%%%%%%%%%%%%%%%%%%%%%%%%%%%%%%%%%%%%%
\section{Conclusions}
\label{conclusions}

In this study we explored the cosmic equation of state across the QCD transition for a range of baryon and lepton asymmetry scenarios, extending our previous work \cite{Gonin:2025uvc} to include non-vanishing chemical potentials and to show the complete thermal history in the presence of X17 bosons. Using a microscopic QCD model \cite{Blaschke:2023pqd} combined with state-of-the-art lattice QCD susceptibilities up to 4th order \cite{Formaggio:2025nde}, we computed cosmic trajectories in the nuclear matter phase diagram and their thermodynamic consequences over a temperature range up to $T=10\rm~GeV$ by including charm quark corrections. We motivate the computation of susceptibilities of heavy quarks in a cosmological context. The extension to $T<10\rm~MeV$ is also of crucial importance, as this epoch corresponds to the formation of SMBHs. A future study is already planned.

Using two simplistic models with high baryonic chemical potential but without LAU, we clarified the role of the baryon and lepton chemical potentials on the thermodynamics, finding the possibility of a stiff universe pre-QCD transition that has not been considered so far. Furthermore, the comparison with realistic models highlights the importance of lepton asymmetry on the EoS.
Unlike the previous Taylor expansion and lattice-QCD-based approaches, which rely on distinct hadronic, quark-gluon and non-interacting quark gas regimes, the microscopic model provides continuous trajectories across the transition, offering a smoother and more consistent description of the thermodynamics. We find that the baryon chemical potential $\mu_B$ alone does not capture the full picture of the primordial plasma at the QCD transition: lepton flavor asymmetries, through electric charge conservation, generate comparable or larger chemical potentials in the QCD sector and significantly mitigate the EoS softening. The resulting PBH mass distributions show distinct features — in particular a pre-QCD stiffening and a modified QCD peak — that differ qualitatively from the standard scenario. The recent LVK sub-solar mass candidate \cite{LISACosmologyWorkingGroup:2023njw,2025GCN.42724....1S} and its compatibility with lepton-asymmetry-driven PBH scenarios \cite{2026arXiv260106024C} motivate a deeper quantitative comparison, which we plan to address in a future study including a more systematic exploration of the lepton asymmetry parameter space and a tighter confrontation with gravitational wave observations.

%%%%%%%%%%%%%%%%%%%%%%%%%%%%%%%%%%%%%%%%%%%%%%%%%%%%%%%%%
%%%%%%%%%%%%%%%%%%%%%%%%%%%%%%%%%%%%%%%%%%%%%%%%%%%%%%%%%

%\authorcontributions{For research articles with several authors, a short paragraph specifying their individual contributions must be provided. The following statements should be used ``Conceptualization, X.X. and Y.Y.; methodology, X.X.; software, X.X.; validation, X.X., Y.Y. and Z.Z.; formal analysis, X.X.; investigation, X.X.; resources, X.X.; data curation, X.X.; writing---original draft preparation, X.X.; writing---review and editing, X.X.; visualization, X.X.; supervision, X.X.; project administration, X.X.; funding acquisition, Y.Y. All authors have read and agreed to the published version of the manuscript.'', please turn to the  \href{http://img.mdpi.org/data/contributor-role-instruction.pdf}{CRediT taxonomy} for the term explanation. Authorship must be limited to those who have contributed substantially to the work~reported.}

\funding{
The research of D.B. and O.I. was supported by the NCN grant No. 2021/43/P/ST2/03319.  
}

\acknowledgments{
We thank Alberto Magaraggia for the code plotting Fig.~\ref{fig:pbh_proba_density}, Lorenzo Formaggio, Francesco Di Clemente, Dominik Schwarz, Julien Froustey, Albert Escriv\'{a} and Florian K\"uhnel for the fruitful discussions.
 }

%%%%%%%%%%%%%%%%%%%%%%%%%%%%%%%%%%%%%%%%%%%%%

%%%%%%%%%%%%%%%%%%%%%%%%%%%%%%%%%%%%%%%%%%
%\isPreprints{}{% This command is only used for ``preprints''.
%\begin{adjustwidth}{-\extralength}{0cm}

\reftitle{References}


\bibliography{PBH_Heraeus}

\PublishersNote{}
%\isPreprints{}{% This command is only used for ``preprints''.
%\end{adjustwidth}

\appendix
\section{Base thermodynamics}\label{app:BaseThermo}
\begin{figure}
    \centering
    \includegraphics[width=\linewidth]{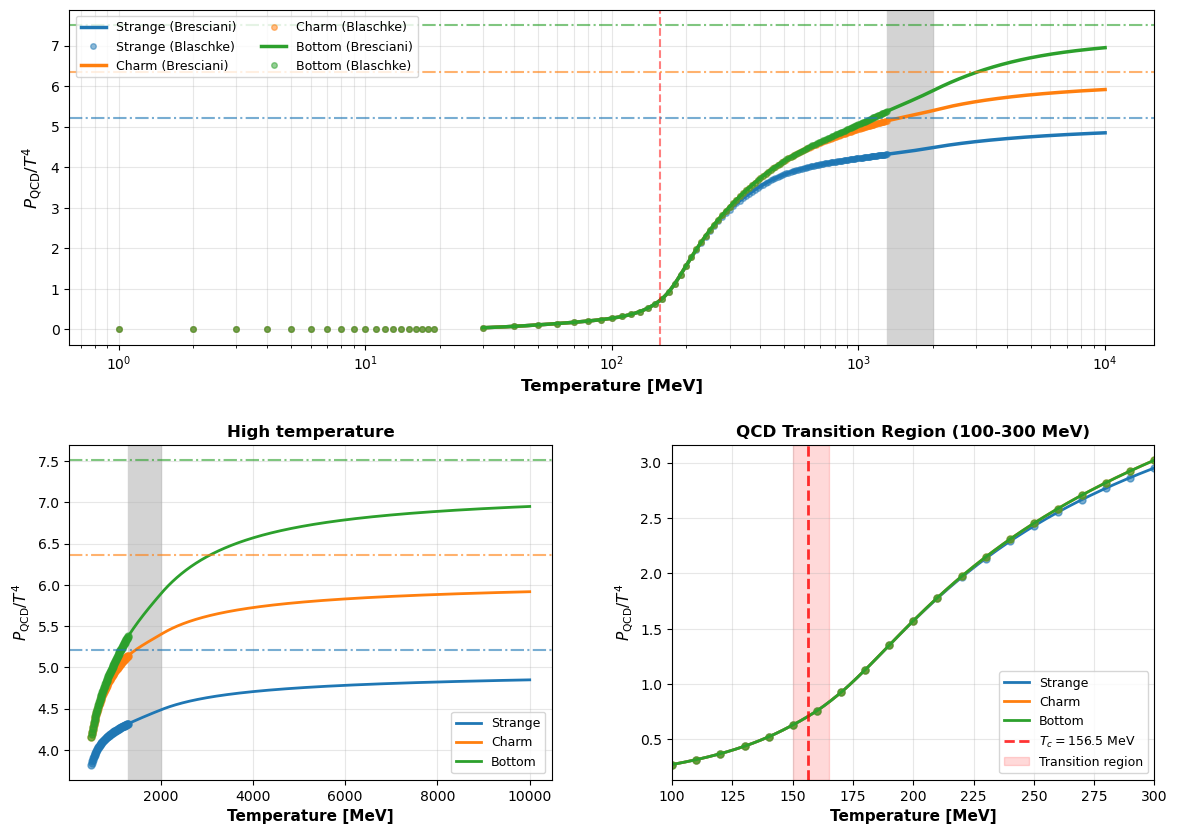}
    \caption{Base QCD pressure for different numbers of quark flavors; the legend shows the heaviest quark of each configuration. Dots represent the base pressure from Ref.~\cite{Blaschke:2023pqd} down to $T=1$ MeV. The dash-dotted lines correspond to the SB limits, while the continuous lines show the data used for cosmic trajectory calculations; for $T>2000 \rm MeV$ we used the polynomial extrapolation from Ref.~\cite{Bresciani:2025vxw}, and the microscopical data extends up to $T=1300 \rm ~MeV$. We rely on PCHIP interpolation, marked with the gray shaded region.}
    \label{fig:base_pressure}
\end{figure}

In Fig.~\ref{fig:base_pressure} we show different base pressures from the microscopical model and Bresciani in blue, and in orange and green the tree-level evaluation using Eq.~\ref{eq:charmCorrection}. We can see how little impact the addition of heavy quarks has on the QCD transition at $T\sim156 \rm~MeV$. 
The EW scale is not considered here, as it could be the baryogenesis and/or leptogenesis era, making the evolution of the thermodynamic quantities very model dependent \cite{Bodeker:2020ghk}. The charge conservation equations \ref{eq:conservation_equations} would not apply with the left-hand side as constant values.
The orange curve corresponds to the base data used to compute the cosmic trajectories (Sec.~\ref{sec:method} and \ref{sec:results}).
The green curve is the one used in Figs.~\ref{fig:EoS_std} and \ref{fig:PBH_spectrum_SM_X17}.

\section{More cosmic trajectories}
\begin{figure}
    \centering
    \includegraphics[width=0.9\linewidth]{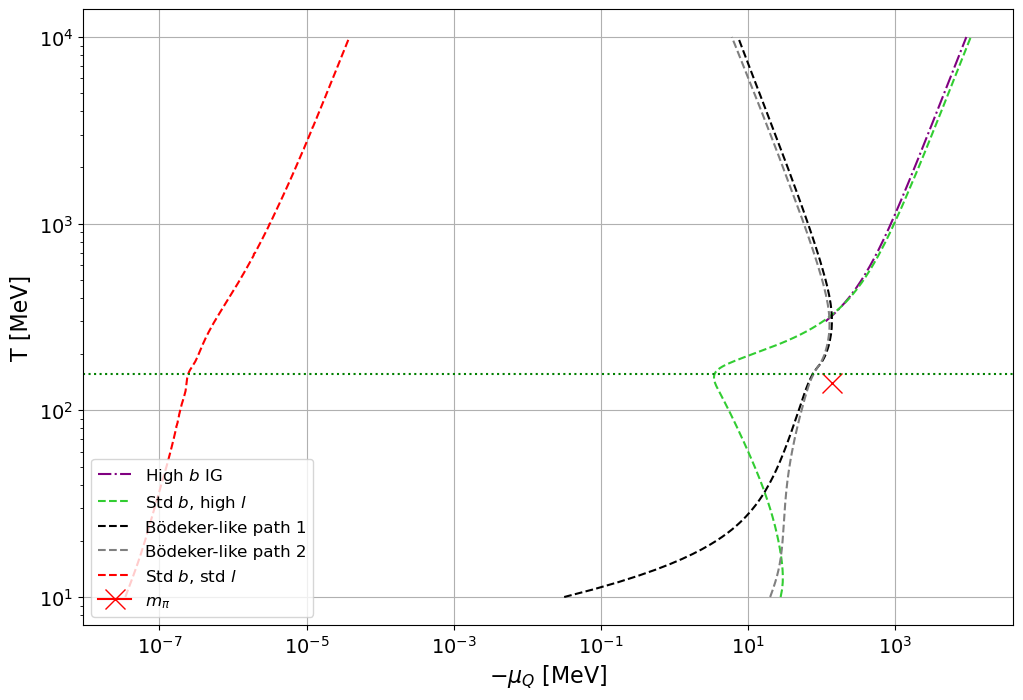}
    \caption{Trajectories in the $T \rm vs$ $\mu_Q$ plane. The trajectories with $\mu_B/T$ fixed are not shown as $\mu_Q$ is ignored. If a trajectory were on the right on $\mu_\pi$ it would indicate the formation of a pion condensate.}
    \label{fig:Tvsmu_Q}
\end{figure}

\begin{figure}
    \centering
    \includegraphics[width=0.9\linewidth]{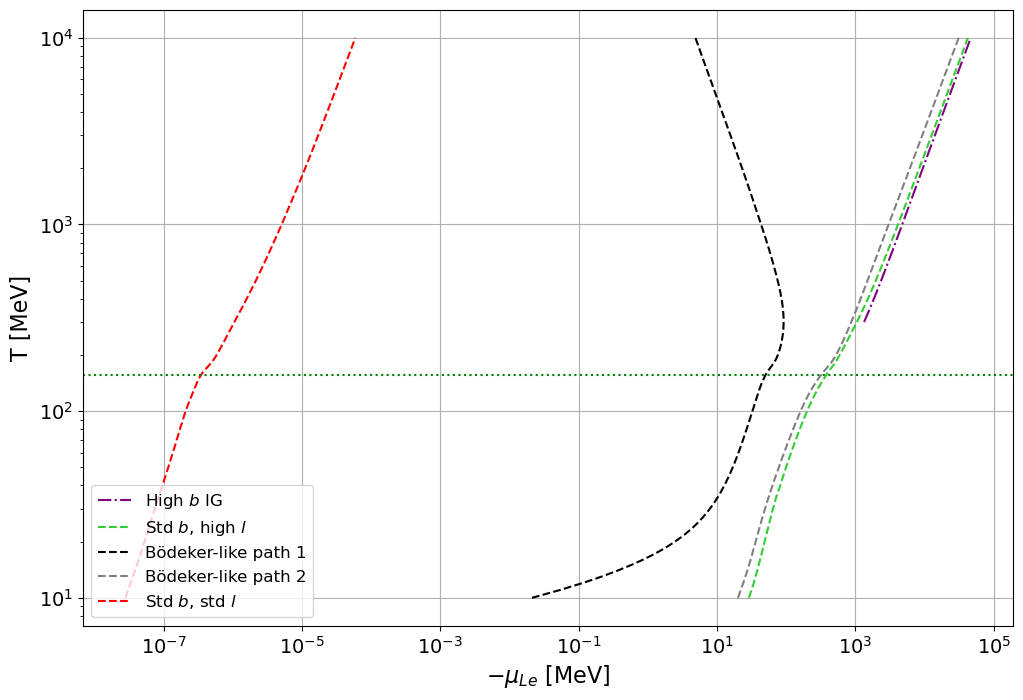}
    \caption{Trajectories in the $T \rm vs$ $\mu_{L_e}$ plane.}
    \label{fig:Tvsmu_Le}
\end{figure}

In Sec.~\ref{sec:EoS} we presented the equations to solve and the induced cosmic trajectories in the $T \rm vs$ $\mu_B$ plane. The system of equations being solved in the direction of the 5 chemical potentials, we show in this Appendix the cosmic trajectories in the other directions. Neutrinos $\nu_\alpha$ and their associated charged leptons $\alpha$ are linked according to \ref{eq:part_chem_pot:c}. Comparing the magnitude of $\mu_Q$ Fig.~\ref{fig:Tvsmu_Le} and $\mu_{L_e}$ Fig.~\ref{fig:Tvsmu_Q} it becomes obvious that $\mu_e$ is small. The same analysis can be done for muon and tau leptons. In the 'B\"odeker-like 2', $\mu_e$ gains a significant value at large temperature, while for 'Std $b$, high $l$' the bounce at the QCD transition lets $\mu_e$ gain value before vanishing at low temperature.

We also observe the difference in behavior between the 'B\"odeker-like 1' and 'B\"odeker-like 2' in the $T \rm vs$ $\mu_{L_e}$, where the former don't have asymmetry in the electronic sector and the latter does, which show a similar behaviour to 'Std $b$, high $l$'.

\begin{figure}
    \centering
    \includegraphics[width=0.9\linewidth]{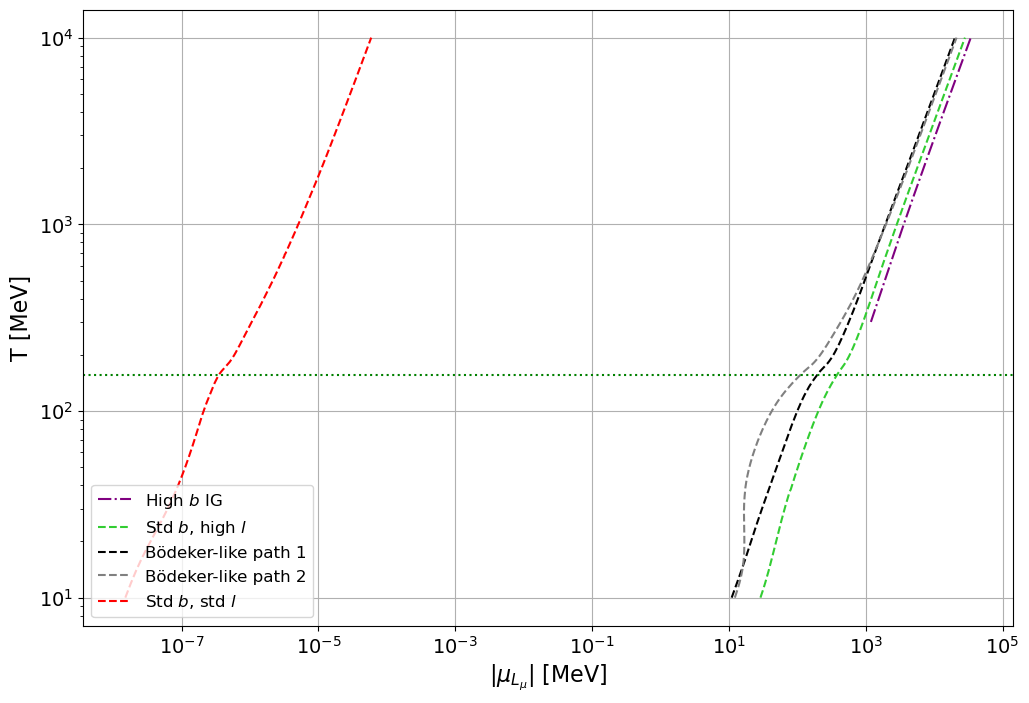}
    \caption{Trajectories in the $T \rm vs$  $\mu_{L_\mu}$ plane. 'High $b$ IG', 'Std $b$, high $l$' and 'B\"odeker-like 2' exhibit positive chemical potential as the muon asymmetry is positive; for 'B\"odeker-like 1' and 'Std $b$, std $l$' we show $-\mu_{L_\mu}$. }
    \label{fig:Tvsmu_Lmu}
\end{figure}

\begin{figure}
    \centering
    \includegraphics[width=0.9\linewidth]{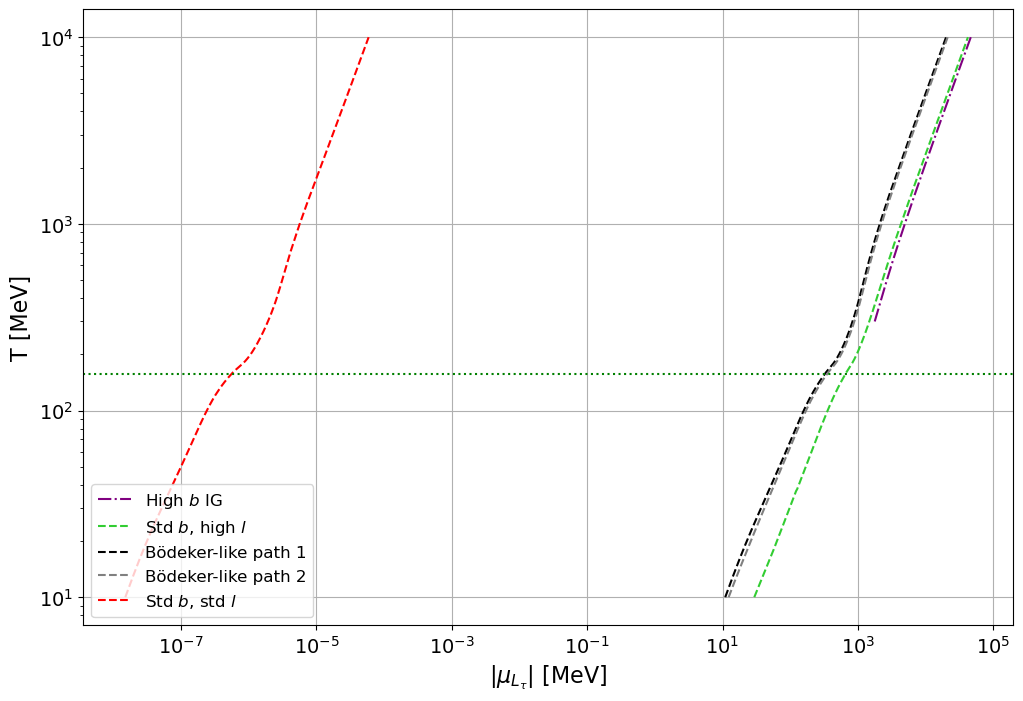}
    \caption{Trajectories in the $T \rm vs$  $\mu_{L_\tau}$ plane. 'B\"odeker-like 1' and 'B\"odeker-like 2' exhibit positive chemical potential as the tau asymmetry is positive; for the remaining model we show $-\mu_{L_\tau}$.}
    \label{fig:Tvsmu_Ltau}
\end{figure}

\end{document}